\documentstyle[prd,aps]{revtex}
\input epsf.tex
\def\DESepsf(#1 width #2){\epsfxsize=#2 \epsfbox{#1}}
\begin{document}
\draft
\twocolumn[\hsize\textwidth\columnwidth\hsize\csname
@twocolumnfalse\endcsname


\title{{\Large\bf Bimaximal Neutrino Mixing From a Local
$SU(2)$ Horizontal Symmetry }}
\author{{\bf R. Kuchimanchi} and {\bf R.N. Mohapatra}}

\address{Department of Physics, University of Maryland, College Park, MD-
20742}
\date{July, 2002}

\maketitle

\begin{abstract}
Freedom from global anomalies in the presence of a local $SU(2)_H$
horizontal symmetry under which right handed charged leptons transform
nontrivially requires that there be at least two right handed
neutrinos with masses of order of the horizontal symmetry breaking scale. 
If a third right handed neutrino is introduced to satisfy quark lepton
symmetry, it is unprotected by the horizontal symmetry, becomes
superheavy with a Planck scale mass and decouples from lower energy
physics. The resulting seesaw mechanism with two right handed neutrinos 
in combination with the horizontal symmetry leads naturally to a near
bimaximal pattern for the neutrino mixing with an inverted mass hierarchy
and is compatible with all data. It predicts a correlation between the solar
mixing angle and $U_{e3}$, that is testable in the proposed long
baseline experiments.
\end{abstract}

\pacs{UMD-PP-02-60}
\vskip2pc]

\section{Introduction}
It is well known that the standard model predicts massless neutrinos due
to the presence of an exact global $B-L$ symmetry. Since it is generally
believed that global symmetries are broken by nonperturbative
Planck scale effects, nonvanishing neutrino
masses\cite{weinberg} of order $m_{\nu}\simeq
\frac{v^2_{wk}}{M_{P\ell}}\simeq 10^{-5}$ eV can arise within the
standard model framework from gravitational effects. Neutrino masses of
such magnitudes are
however much too small to account for observed neutrino
oscillations. In fact atmospheric neutrino
data requires that there must at least be one neutrino with
considerably larger mass (of order $0.05$ eV), implying that 
the scale of new physics responsible for breaking the $B-L$ symmetry of
the standard model must be much lower scale than the Planck scale ($\leq
10^{-4} M_{P\ell}$).
In the context of the seesaw model for understanding the small neutrino
masses\cite{seesaw} where heavy right handed neutrinos are responsible
for breaking $B-L$ symmetry and hence for small neutrino masses,
atmospheric neutrino data implies that their typical
masses should be around $\leq 10^{-4} M_{P\ell}$. 
Thus the right handed neutrino
masses must be protected by additional symmetries. In the SO(10) or
left-right models which provide the conventional venues for implementing
the seesaw mechanism, the relevant symmetry is the local $B-L$ symmetry.
Requiring that an extension of the standard model respect the
$B-L$ symmetry implies that there must three right handed neutrinos
$(\nu_{eR}, \nu_{\mu R}, \nu_{\tau R})$ to
cancel the gauge anomalies. All the $\nu_R$'s are expected to have masses
much lower than the Planck scale and of order of the $B-L$
symmetry breaking scale.

In this note we explore an alternative approach to keeping the right
handed neutrinos lighter than the Planck scale. Instead of the local
$B-L$ symmetry (or perhaps
in addition to it) we propose that there be a local
 $SU(2)_H$ horizontal symmetry at the seesaw scale under which the
right handed charged leptons transform nontrivially. There are then two
possibilities which are of interest for our considerations: (i) $SU(2)_H$
acts on both
right handed quarks and leptons or (ii) it acts only on leptons
from considerations of gauge anomalies. Freedom from
global Witten anomaly on the other hand requires that in both cases
 there must be two right handed neutrinos that trasform as a doublet
of $SU(2)_H$. The local $SU(2)_H$ symmetry then implies that the masses of
those two right handed neutrinos are protected and must be at the
scale of $SU(2)_H$ breaking. If there is a the third right handed neutrino
for reasons of quark lepton symmetry, then it will acquire a mass of order
of the Planck or string scale and decouple from neutrino physics at
lower energies. This
therefore provides a physically distinct way of implementing the seesaw
mechanism. It leads to a neutrino mass pattern that is inverted. 
Furthermore, it has the interesting property that it
leads to the near bimaximal mixing pattern in a more natural manner than
the $B-L$ seesaw approach. The $SU(2)_H$ symmetry plays a crucial role in
generating the near bimaximal pattern. This mixing pattern seems
to be favored by the detailed analyses of present solar and
atmospheric neutrino data\cite{bahcall}.

The fact that dominance of two right handed
neutrinos in a seesaw model under certain circumstances can lead to
 bimaximal mixing needs was noted in a different context in \cite{lavoura}. 
$SU(2)_H$ horizontal symmetry not only provides a rational for this
hypothesis but also naturally leads to a near bimaximal mixing. We will
also see that it provides a natural understanding of why $\Delta
m^2_{\odot}/\Delta m^2_A \ll 1$.

There are two possible scenarios that arise in these models depending on
how charged leptons are aligned. We discuss only one of them in detail in
this paper. This model has the following predictions.

(i) The sign of the $\Delta m^2_A$ is opposite to the
case of normal hierarchy for neutrinos.

(ii) There is a correlation between the value of
$U_{e3}$ and the solar mixing angle $sin^22\theta_{\odot}$ as in a class
of softly broken $L_e-L_{\mu}-L_{\tau}$ models\cite{babu} i.e. a lower
$sin^22\theta_{\odot}$ requires a higher $U_{e3}$. This correlation is
testable in future long baseline experiments such as NUMI Offaxis
proposal\cite{para} or proposed JHF facilities.

(iii) The effective neutrino mass
$m_{\beta\beta}$ that can be measured in neutrinoless double beta decay
experiments is related to the atmospheric mass difference squared $\Delta
m^2_{A}$ and
$U_{e3}$ i.e. $m_{\beta\beta} \simeq 2 U_{e3} \cdot \sqrt{2\Delta m^2_A}$.

While in this paper, we have only considered the $SU(2)_H$ symmetry to act
on the leptonic sector, aesthetic reasons suggest that the $SU(2)_H$ could
act simultaneously on quarks and leptons. We note that in a model of this
type, our conclusion regarding only two RH neutrinos at the $SU(2)_H$
scale still remains valid as long as the right handed charged leptons
transform as a doublet. The details implications of this model are
currently under investigation.

 \section{The $SU(2)_H$ model}
Horizontal symmetries have often been invoked to understand family
replication and flavor
structure of the quarks and leptons in extensions of the standard
models. They can
either be  $U(1)_H$, $SU(2)_H$ or $SU(3)_H$ type. We will consider
$SU(2)_H$ type models\cite{hor}, which seem to have interesting
consequences for neutrino mixings.

Gauge anomaly constraints can be satisfied in four distinct ways for
an $SU(3)_c\times SU(2)_L\times U(1)_Y\times SU(2)_H\equiv G_{STD}\times
SU(2)_H$ model if we consider only the known fermions: (i) only left
handed fermions
of the standard model transform as $SU(2)_H$
doublets; (ii) only quarks (both left and right handed) transform as the
$SU(2)_H$ doublets; (iii) only
right handed fermions (both quarks and leptons) transform as $SU(2)_H$
doublets and (iv) only leptons transform as doublets. 
In cases (iii) and (iv), freedom from global $SU(2)_H$ Witten anomaly
implies that there must be at least two right handed neutrinos (we will
call them $(\nu_{eR}, \nu_{\mu,R})$), transforming as a doublet of
$SU(2)_H$.
We show below that this model leads naturally to the near bimaximal
pattern for neutrino mixings.

We will focus in this paper on the possibility that only leptons
transform under $SU(2)_{H}$ and quarks are singlets. This is the minimal
model with an $SU(2)_{H}$. Our considerations can be extended to the other
case with small modifications.
We give below in Table I the assignment of fermions and Higgs bosons
under the gauge group $ G_{STD} \times SU(2)_H$

\begin{center}
{\bf Table I}
\end{center}

\begin{center}

\begin{tabular}{|c||c|}\hline
Particles &  $G_{STD}\times SU(2)_H$\\
    & Quantum numbers \\ \hline\hline
$\Psi \equiv (\psi_e, \psi_{\mu})$ & (1,2,-1,2) \\ \hline
$\psi_{\tau}$  &  (1,2,-1,1) \\ \hline
$E_R \equiv (e_R, \mu_R)$ & (1,1,-2, 2) \\ \hline
$\tau_R$ & (1,1,-2, 1)\\ \hline
$N_R\equiv (\nu_{eR}, \nu_{\mu R})$ & (1,1,0,2) \\ \hline
$\nu_{\tau R}$ & (1,1,0,1) \\ \hline
$\Phi\equiv \left(\begin{array}{cc} \phi^0_{1} & \phi^0_2 \\ 
\phi^-_1 & \phi^-_2\end{array}\right) $ & (1, 2, -1, 2) \\ \hline
$\chi_H$ & (1,1,0,2) \\ \hline
$\phi_0$ & (1,2,-1,1)\\ \hline
$\Delta_H$ & (1,1,0,3)\\ \hline\hline
\end{tabular}

\end{center}
\bigskip

\noindent{\bf Table caption}: Representation content of the
various fields in the model under the gauge group $G_{STD}\times SU(2)_H$.

\bigskip
Here $\psi_{e,\mu,\tau}$ denote the left handed lepton doublets.
We arrange the Higgs potential in such a way that the $SU(2)_H$ symmetry
is broken by $<\chi_1>= v_{H1}; <\chi_2>=v_{H2}$ and $<\Delta_{H,3}>=
v'_H$, where $v_H,v'_H \gg v_{wk}$. Note that we have used the $SU(2)_H$
symmetry to align the $\Delta_H$ vev along the $I_{H,3}$ direction. At the
weak scale,
all the neutral components of the fields $\Phi$ and $\phi_0$ acquire
nonzero vev's and break the standard model symmetry down to $SU(3)_c\times
U(1)_{em}$. We denote these vev's as follows: $<\phi^0_0> =\kappa_0$;
$<\phi^0_1>= \kappa_1$ and $<\phi^0_2>=\kappa_2$. Clearly $\kappa$'s
have values in few to 100 GeV range. As we discuss later, we expect a
hierarchy between the two vevs $\kappa_1$ and $\kappa_2$, which is
important in our discussion of neutrino mixings.

Note that $<\Delta_H>\neq 0$ breaks
the $SU(2)_H$ group down to the $U(1)_{L_e-L_{\mu}}$ group which is
further broken down by the $\chi_H$ vev. Since the renormalizable
Yukawa interactions do not involve the $\chi_H$ field, this symmetry
($L_e-L_{\mu}$) is also reflected in the right handed neutrino mass
matrix.

To study the pattern of neutrino masses and mixings, let us first
note that a bare mass for the $\nu_{\tau, R}$ field is allowed at the 
tree level unconstrained by any symmetries. This mass can therefore
be arbitrarily large and $\nu_{\tau, R}$ will decouple from
the low energy spectrum. We will work in this limit of decoupled
$\nu_{\tau R}$ and write
down the gauge invariant Yukawa couplings involving the remaining leptonic
fields.
\begin{eqnarray}
{\cal L}_Y~=~h_1\bar{\psi}_{\tau}{\Phi}N_R
+h_0Tr(\bar{\Psi}{\phi}_0N^T_R)\\ \nonumber
-if N^T_R\tau_2{\bf \tau \cdot \Delta_H}N_R \\ \nonumber
h'_1 Tr(\bar{\Psi}\tilde{\Phi}_2\tau_2) \tau_R
+h'_2\bar{\Psi}\tilde{\phi}_0E^T_R\\
\nonumber
h'_3 \bar{\psi}_{\tau}\tilde{\phi_0}\tau_R +
h'_4\bar{\psi}_{\tau}\tilde{\Phi}E_R 
+ h.c.
\end{eqnarray}
where $\tilde{\Phi}=\tau_2 \Phi^*\tau_{2 H}$.
Since $<\Delta^0_H>= v'_H$ directly leads to the $L_e-L_{\mu}$ invariant
  $\nu_{eR}-\nu_{\mu R}$ mass matrix at the seesaw scale.
The $\chi_H$ vev contributes to this mass matrix only through
nonrenormalizable operators and its contributions are therefore
negligible. Similarly there will also be some small contributions from the
$\nu_{\tau R}$ sector if we did not decouple it completely. We ignore
these small contributions in our analysis. 
As we will see below, this feature of the right handed neutrino sector is
crucial to the light neutrino mass matrix that
leads in the zeroth order to bimaximal mixing. To see this, let us write
down the $5\times 5$ seesaw matrix for neutrinos:
\begin{eqnarray}
M_{\nu_L,\nu_R}~=~\left(\begin{array}{ccccc} 0 & 0 & 0 & h_0\kappa_0 & 0\\
0 & 0 & 0 & 0 & h_0\kappa_0\\ 0 & 0 & 0 & h_1\kappa_1 & h_1 \kappa_2 \\
h_0\kappa_0 & 0 & h_1\kappa_1 & 0 & fv'_H \\ 0 & h_0\kappa_0 & h_1\kappa_2
& fv'_H & 0 \end{array} \right)
\end{eqnarray}
After seesaw diagonalization, it leads to the light neutrino mass matrix
of the form:
\begin{eqnarray}
{\cal M}_{\nu}~=~-M_D M^{-1}_R M^T_D
\end{eqnarray}
where $M_D~=~\left(\begin{array}{cc} h_0\kappa_0 & 0 \\ 0 & h_0\kappa_0\\
h_1\kappa_1 & h_1\kappa_2
\end{array}\right)$; $M^{-1}_R~=~\frac{1}{fv'_H}\left(\begin{array}{cc} 0
&
1\\1 & 0 \end{array}\right)$. The resulting light Majorana neutrino mass
matrix ${\cal M}_{\nu}$ is given by:
\begin{eqnarray}
{\cal M}_{\nu}~=~-\frac{1}{fv'_H}\left(\begin{array}{ccc} 0 &
(h_0\kappa_0)^2 & h_0h_1\kappa_0\kappa_2\\ (h_0\kappa_0)^2 & 0 &
h_0h_1\kappa_0\kappa_1 \\ h_0h_1\kappa_0\kappa_2 & h_0h_1 \kappa_0\kappa_1
& 2h^2_1\kappa_1\kappa_2 \end{array}\right)
\end{eqnarray}
To get the physical neutrino mixings, we also need the charged lepton mass
matrix defined by $\bar{\psi}_L {\cal M}_\ell \psi_R$. This is given in
our model by:
\begin{eqnarray}
{\cal M}_{\ell}~=~\left(\begin{array}{ccc} h'_2\kappa_0 & 0 &h'_1\kappa_1
\\ 0 & h'_2\kappa_0 & h'_1\kappa_2 \\ -h'_4\kappa_2 & h'_4\kappa_1 &
h'_3\kappa_0 \end{array}\right)
\end{eqnarray}
Note that in the limit of $\kappa_1 =0$, the neutrino mass matrix,
have exact $(L_e-L_{\mu}-L_{\tau})$ symmetry\cite{petcov} where the
charged lepton mass
matrix breaks this symmetry.

The invariance of the neutrino mass matrix under $L_e-L_{\mu}-L_{\tau}$
symmetry in the limit of $\kappa_1 = 0$ and $\kappa_2\neq 0$ is an
important issue since it leads to a neutrino mixing pattern which is 
very close to what is apparently observed in neutrino oscillation
experiments and as such has been widely considered in the
context of gauge models\cite{emutau,babu}. Further in this limit, we have
$\Delta m^2_{\odot}=0$. To generate solar neutrino
oscillation however, a small but nonzero $\Delta m^2_{\odot}$ is
needed. This happens as soon as a small
$\kappa_1\ll \kappa_2$ is turned on. It turns out that $(\Delta
m^2_{\odot}/\Delta m^2_A)$ is proportional to $\kappa_1/\kappa_2$.
Thus the observed smallness of  $(\Delta
m^2_{\odot}/\Delta m^2_A)$ is protected by a symmetry
if $\kappa_1/\kappa_2$ is.
We show in the next section that this is indeed the case
 in our model due to the presence of a discrete $Z_2$
symmetry in combination with the horizontal symmetry. This makes the 
neutrino mixing pattern with softly broken $L_e-L_{\mu}-L_{\tau}$ in the
neutrino sector that is responsible for solar neutrino oscillation a
natural consequence of our model. The third neutrino has zero mass in our
model.

\section{Naturalness of softly broken $L_e-L_{\mu}-L_{\tau}$ symmetry}
 To demonstrate that a softly broken $L_e-L_{\mu}-L_{\tau}$ for leptons
arises in a natural manner in our model, we have to show that
$\kappa_1\ll \kappa_2$ is natural and does not receive infinite radiative
corrections. For this purpose, we
write the Higgs potential of the theory as a sum of two parts
\begin{eqnarray}
V(\Phi, \phi_0, \Delta_H, \chi_H)~=~ V_0 + \mu_0Tr(\Phi{\bf \tau\cdot
\Delta_H}\Phi^{\dagger})\\ \nonumber
 + \phi^{\dagger}_0\Phi (\mu'\chi_H+\tilde{\mu}'\tilde{\chi}_H) + M^2
Tr{\Phi^{\dagger}\Phi}\\ \nonumber
+\mu''\chi^{\dagger}_H{\bf \tau\cdot \Delta_H}\chi_H + h.c.
\end{eqnarray}
where $V_0$ contains the standard $\phi^{\dagger}\phi$ type terms
involving all the Higgs fields as well terms that are not relevant
to the discussion of the $\kappa_1/\kappa_2$ and $\tilde{\chi}_H$ is
defined as $\tau_2\chi^*_H$. Note
that in the limit when the parameter $\mu',\tilde{\mu}'=0$, the theory
has $Z_2$
symmetry under which only $\chi_H$ field changes sign and all other
fields remain unchanged. The $\mu'$ and $\tilde{\mu}'$ break this symmetry
softly and
can therefore be chosen to be small compared to $\mu_0$. We do not include
in the potential the term $\phi^{\dagger}_0\Phi\Delta_H\chi_H$ term which
forbidden by $Z_2$ symmetry which could have broken the $Z_2$ symmetry in
a ``hard'' way. It is then
easy to show that all radiative corrections to $\mu'$ and $\tilde{\mu}'$ 
are proportional to $\mu'$  and $\tilde{\mu}'$. The effects of $\mu'$ and
$\tilde{\mu}'$ are similar; so henceforth we will only refer to $\mu'$.

Let us now show that $\kappa_1$ owes its origin to $\mu'$.
For this purpose, we need to discuss the breaking of the horizontal
symmetry in some more detail.

We choose the mass terms for the $\Delta_H$  and $\chi_H$ fields to have 
negative sign
so that at the minimum of the potential they acquire vevs $v'_H$ and
$v_H$ respectively, breaking the $SU(2)_H$ symmetry down to
$U(1)_{L_e-L_{\mu}}$ and subsequently no horizontal symmetry.
Turning to their effect on the vevs of the standard model doublets
$(\phi_{1,2}$ in $\Phi$ ($\Phi\equiv (\phi_1, \phi_2)$), 
we see that for $M^2\geq 0$,
$<\Phi>=0$ in the absence of horizontal symmetry breaking. Once horizontal
symmetry breaking by $\Delta_H$ is turned on i.e. $v'_H\neq 0$, 
for $\mu'=0$, the $\mu_0$ term gives
contributions to the masses of the two standard model doublets in $\Phi$
with opposite sign so that one has $m^2_{\phi_1} = M^2+\mu_ov'_H$ whereas
$m^2_{\phi_2}=M^2 - \mu_0v'_H$. We can choose $\mu_0v'_H$
such that $m^2_{\phi_2}$ becomes negative and of order of the electroweak
scale leading to $\kappa_2\neq 0$; but at the same time since
$m^2_{\phi_1}\geq 0$, it keeps
$\kappa_1 = 0$. It is clear from examination of the Higgs
potential that $\kappa_1$ remains zero as long as $\mu'=0$ (i.e. in
the limit of exact $Z_2$ symmetry). This is also
preserved by radiative corrections due to the presence of the discrete
symmetry, $Z_2$.

 Once $\mu'\neq 0$, the $\phi^{\dagger}_0\Phi\chi_H$ term in Eq. (9) 
induces a nonzero vev for the electrically neutral member of the doublet
$\phi_1$ giving $\kappa_1 \simeq
\mu' v_H \kappa_0/2M^2$. Since $\mu'$ is a soft symmetry breaking
parameter, we can choose it appropriately small to obtain $\kappa_1 \ll
\kappa_2$. Thus the presence of the horizontal symmetry is crucial to
obtaining the desired pattern for the neutrino mass matrix in our model.

As noted there are two horizontal symmetry breaking scales $v'_H$ and
$v_H$ in this model. The first vev $v'_H$ corresponds to the seesaw scale
and is therefore determined by the
neutrino masses. For instance, if we choose the Dirac masses of the
neutrinos to be $\sim$ GeV, we get $v'_H \simeq 10^{11}$ GeV. In general
this scale is likely to be anywhere between $10^{15}$ GeV to $10^{9}$ GeV
for Dirac masses between $0.1$ GeV to 100 GeV. 
On the other
hand the $\chi_H$ vev $v_H$ in principle can be much lower. The $\chi_H$
vev gives mass to the horizontal gauge boson corresponding to the diagonal
generator of $SU(2)_H$, which couples to both electrons and muons. To fit
present
neutral current observations involving the electrons and muons, one must
have
$\chi_H$ vev at least a few TeV. However to generate $\kappa_1\simeq
\kappa_2/10$ (see the rational for this choice in the next section), we
must have $v_H \sim v'_H$.

\section{Charged lepton spectrum}
In order to discuss the physical neutrino mixings, we need to work in a
basis where the charged leptons are diagonal. There are two ways to get
the right charged lepton mass hierarchy in our model:
(i) First way is to choose $\kappa_1 \ll \kappa_2 \simeq \kappa_0$ and
$h'_2\kappa_0\gg m_e$ and 
(ii) a second way is where $m_e = h'_2\kappa_0$. We consider only the
first case here. In
this case for muon and tau lepton masses, we get roughly
$m_{\tau}\simeq\sqrt{( h'_3\kappa_0)^2 + (h'_4\kappa_2)^2}$ and
$m_{\mu}\simeq h'_2\kappa_0$ and for
electron, we get
$ m_e \simeq  h'_2h'_4\kappa_2\kappa_0/m_{\tau}$.
 We will see in the next section that we get
 $U_{e3} \simeq h'_4\kappa_2h'_2\kappa_0/m^2_{\tau}\sqrt{2}$, which we
will demand to be of order $0.1$. All these constraints can be satisfied
since we have five free parameters in $M_\ell$; however, they must satisfy
certain constraints e.g. we must have $h'_4 \gg h'_1$ in addition to th
constraints implied by the mass relations given above.

It is clear that we need a certain degree of fine tuning in the charged
lepton sector.
 This fine tuning is however needed only in this minimal
version of the horizontal model. For instance if there are two sets of
Higgs doublets, one coupling to charged leptons and another to neutrinos,
as would be the case in a supersymmetrized version of the model,
the charged leptons and the neutrinos get their masses from different
Higgs multiplets. As a result, one can get a realistic charged lepton
spectrum without fine tuning while preserving other consequences of
horizontal symmetry such as the connection between the $U_{e3}$ and solar
mixing angle. The important point is that the presence of a horizontal
symmetry leads to a nonvanishing $U_{e3}$ which makes the model
phenomenologically interesting.

\section{Solar neutrino mixing angle and $U_{e3}$ connection}
We now turn to the discussion of neutrino mixings in our model.
As already noted, in the limit of $\kappa_1 =0$, the neutroino mass matrix
has exact $(L_e-L_{\mu}-L_{\tau})$ symmetry. 
The neutrino mixing matrix in this limit has the form:
\begin{eqnarray}
U^{\nu}~=~\left(\begin{array}{ccc} \frac{1}{\sqrt{2}} &
-\frac{1}{\sqrt{2}} & 0 \\ \frac{s}{\sqrt{2}} & \frac{s}{\sqrt{2}} & c\\
\frac{c}{\sqrt{2}} & \frac{c}{\sqrt{2}} & -s \end{array}\right)
\end{eqnarray} 
This mixing pattern is quite close to the observed values. Further more,
it corresponds to the inverted mass hierarchy where the solar pair of
neutrinos are heavier. Atmospheric neutrino data imply that $s\simeq
1/\sqrt{2}$. For solar neutrino oscillation, it predicts
$sin^22\theta_{\odot} = 1$. However, since the two heavy eigenstates are
exactly
degenerate, it leads to no solar neutrino oscillation (since  $\Delta
m^2_{\odot}=0$). Furthermore, the third neutrino has zero mass. 
We also note that as soon as $\kappa_1$ is turned on,
we get $\Delta m^2_{\odot} \neq 0$ and there is departure from the
exact bimaximal mixing pattern in Eq. (7). How far the neutrino mixing
matrix can depart from the exact bimaximal form in Eq. (7) will determine
how viable the model is since after the SNO neutral current
results\cite{sno}, combined analyses\cite{bahcall} of all
solar neutrino data\cite{sno,solar} disfavor exact bimaximal mixing.

 We find that to fit
the central value of $\Delta m^2_{\odot}$ required by data, we need to
have $\frac{h^2_1\kappa_1\kappa_2}{fv'_H},
\frac{h_0h_1\kappa_0\kappa_1}{fv'_H}\simeq 10^{-3}$ eV.  
If we assume the Yukawa couplings
to be of order one, this implies
that $\kappa_1/\kappa_2 \simeq \frac{1}{50}$. Since $\kappa_1/\kappa_2$ is
related to $Z_2$ symmetry breaking, we have been able to relate the
smallness of $\Delta m^2_{\odot}/\Delta m^2_A$ to a symmetry breaking. 
Thus in this sense the smallness of $m^2_{\odot}/\Delta m^2_A$ is natural
in our model.

Turning now to the effect of $\kappa_1\neq 0$ on the mixing pattern,
unfortunately due to its smallness, the $U_{\nu}$ matrix remains
practically the same as in Eq. (7). Luckily however, the mixing matrix has
a contribution from the charged lepton sector and from Eq. (5), we see
that even in the limit of $\kappa_1 =0$, $M_{\ell}$ breaks
$L_e-L_{\mu}-L_{\tau}$ symmetry. The analysis therefore becomes very
similar to the first paper in reference\cite{babu}. 

To calculate the contribution of the charged lepton sector to the neutrino
mixing, note that since the charged lepton mass matrix is not
symmetric, it is diagonalized by bi-orthogonal
transformations: $U^{\ell L}M_{\ell} U^{\ell R\dagger}$. The
$U^{\ell L}$ is the matrix relevant for neutrino mixing and is given
by
\begin{eqnarray}
U^{\ell L}=\left(\begin{array}{ccc} c_{\alpha} & 0 & s_{\alpha} \\ 0
& 1 & 0 \\ -s_{\alpha} & 0 & c_{\alpha} \end{array}\right)
\end{eqnarray}
where $tan\alpha \simeq h'_4\kappa_2h'_2\kappa_0/m^2_{\tau}$,
The physical
neutrino mixing matrix ${\bf \Large U}$ can now be written down as:
 $U^{\ell L}U^{\nu}$ where
$U^{\nu}$ diagonalizes the Majorana mass matrix of the neutrinos and is
given above.
\begin{eqnarray}
{\bf \Large U} ~= 
~\left(\begin{array}{ccc}\frac{c_{\alpha}+cs_{\alpha}}{\sqrt{2}}
&\frac{-c_{\alpha}+c s_{\alpha}}{\sqrt{2}} & -s s_{\alpha}
\\\frac{s}{\sqrt{2}} & \frac{s}{\sqrt{2}} & c \\
\frac{c c_{\alpha}-s_{\alpha}}{\sqrt{2}}
&\frac{cc_{\alpha}+s_{\alpha}}{\sqrt{2}}
& s c_{\alpha}\end{array}\right)
\end{eqnarray}
We see that the effective solar neutrino mixing angle becomes
less than its maximal value in the presence of the charged lepton mixing
parameter $\alpha$. We also note that $\alpha$ induces an $U_{e3}\simeq
\frac{\alpha}{\sqrt{2}}$. Present upper limits from
CHOOZ-PALO-VERDE\cite{chooz} experiments imply that $U_{e3}\leq 0.16$,
which translates to a limit on $\alpha \leq 0.2$. 
Using this, we get $\sin^2 2\theta_\odot \simeq 0.9$ or higher. We
also note that the lower $U_{e3}$, the higher the $\sin^2 2\theta_\cdot$
required. Hence the model predicts that $U_{e3}$
should be very close to the present upper reactor bound from the
CHOOZ-PALO-VERDE experiments.

 Note that all the discussions above are done at the seesaw scale. 
There are radiative corrections as we extrapolate down to the weak scale 
arising from charged lepton contributions. It turns out however that they
change the solar
mixing angle slightly over and above that already discussed. It has
recently been suggested\cite{lindner} that in some seesaw models there may
also be high scale
contributions due to different masses of the right handed neutrinos, that
could effect the solar mixing angle. In our model the presence of the
horizontal symmetry precludes such corrections. There is also likely to be
some effect on $sin^22\theta_{\odot}$ if the right handed neutrino is not
totally decoupled.

Another prediction of this model is a value for the
neutrino mass measured in neutrinoless
double beta decay and we find $m_{\beta\beta}\simeq 2\sqrt {2\Delta m^2_A}
U_{e3}$. The maximum value for $m_{\beta\beta}$ is therefore $\simeq
0.007$ eV. This can be probed in proposed double beta experiments such as
GENIUS\cite{genius}. If the recently reported evidence for neutrinoless
double beta decay\cite{hans} by the Heidelberg-Moscow group is
confirmed, this model will be ruled out. 

\section{Conclusions and outlook}
In summary, we have shown that if seesaw mechanism is to be responsible
for the bimaximal mixing pattern for neutrino mixings, then a simple way
to derive it from an extension of the standard model is to postulate the
existence of a local $SU(2)_H$ horizontal symmetry under which
right handed leptons transform nontrivially. 
First this guarantees the existence of two right handed
neutrinos, whose masses are at the scale of horizontal symmetry
breaking. Second, this minimal model via the seesaw mechanism leads to a
near bimaximal mixing pattern suggested by solar and
atmospheric neutrino data. We have found two interesting scenarios but in
this note focus only on one.
It predicts (i) a direct
correlation between the mixing parameter $U_{e3}$ and $\Delta m^2_{\odot}$ 
with the solar mixing
angle $sin^22\theta_{\odot}$ as well as (ii) a negative sign for $\Delta
m^2_A$ 
which can tested in proposed long baseline experiments experiments.
The model has the interesting feature that the smallness of the $\Delta
m^2_{\odot}/\Delta m^2_A$ is related to a discrete symmetry of the model,
much like the smallness of the electron mass is related to the presence of
a chiral symmetry in QED. Better understanding the origin of this discrete
symmetry can therefore perhaps explain why $\Delta m^2_{\odot} \ll \Delta
m^2_A$. It also makes a prediction for neutrinoless double beta decay,
which can be tested.

This model could easily be incorporated into models with local
$B-L$ symmetry. One would then need the horizontal symmetry scale to be
much lower than the $B-L$ symmetry scale.

The model can also be supersymmetrized in a straightforward manner by
promoting all the fields to superfields and duplicating the Higgs fields.
Below the horizontal symmetry breaking scale, one then has the MSSM and
the model preserves coupling constant unification. All features of the
neutrino sector remain unchanged and as noted the charged lepton spectrum
then arises from a new set of Higgs fields and no fine tuning is required
to get charged lepton masses.

As we remarked in the beginning of the paper, one could also work with
a horizontal symmetry that operates on the right handed components of both
quarks and leptons, in which case global anomaly freedom  would again
require the presence of two right handed neutrinos as in the case
discussed in the text. We expect our results for the neutrino mixings
to remain unaltered whereas the Higgs sector may need to be
extended to fit the quark mixing angles. We do not pursue this alternative
here.

The work of R. N. M. is supported by the National Science Foundation Grant
No. PHY-0099544. We thank Luis Lavoura for comments on the first version
of the paper.

\end{document}